\documentclass[aip,apl,reprint,superscriptaddress,showpacs]{revtex4-1}

\usepackage{graphicx}
\usepackage{ulem}
\usepackage{color}

\begin{document}

\title{Enhanced thermoelectric properties by Ir doping of PtSb$_2$  with pyrite structure}

\author{Yoshihiro Nishikubo}
\author{Seiya Nakano}
\author{Kazutaka Kudo}
\author{Minoru Nohara} \email{nohara@science.okayama-u.ac.jp}
\affiliation{Department of Physics, Okayama University, Okayama 700-8530, Japan}
\affiliation{Advanced Low Carbon Technology Research and Development Program (ALCA), Japan Science and Technology Agency (JST), 
Tokyo 102-0076, Japan}

\date{\today}

\begin{abstract}
The effects of Ir doping on the thermoelectric properties of Pt$_{1-x}$Ir$_x$Sb$_2$ ($x$ = 0, 0.01, 0.03, and 0.1) with pyrite structure were studied. 
Measurements of electrical resistivity $\rho$, Seebeck coefficient $S$, and thermal conductivity $\kappa$ were conducted. 
The results showed an abrupt change from semiconducting behavior without Ir ($x$ = 0) to metallic behavior at $x$ = 0.01.
The sample with $x$ = 0.01 exhibited large $S$ and low  $\rho$, resulting in a maximum power factor ($S^2/\rho$) of 
43 $\mu$W/cmK$^2$ at 400 K. 
The peculiar ``pudding mold''-type electronic band dispersion could explain the enhanced thermoelectric properties in the metallic state.
\end{abstract}

\pacs{}

\maketitle

Thermoelectric materials are of considerable practical interest because they could be used to generate electricity directly from waste heat. 
A practical thermoelectric material should have high efficiency, which is represented by the dimensionless figure of merit $ZT = S^2T/\rho\kappa$, where $S$, $T$, $\rho$, and $\kappa$ are the Seebeck coefficient, absolute temperature, electrical resistivity, and thermal conductivity, respectively. 
In addition, the electric power that a thermoelectric material can generate is characterized by the power factor PF = $S^2/\rho$.  
Thus, the goal for current research is the development of materials that exhibit both a large Seebeck coefficient $S$ and low (metallic) electrical resistivity $\rho$.

However, most common materials exhibit a trade-off in $S$ and $\rho$.
Specifically, a large Seebeck coefficient $S$ is generally associated with semiconducting behavior due to the large asymmetry in the velocity of charge carriers about the chemical potential when the chemical potential is located near the edge of the conduction or valence band. 
Moreover, low electrical resistivity $\rho$ ($= 1/ne\mu$) is associated with metallic behavior, due to the large carrier density $n$ (i.e., large Fermi surface) and/or large carrier mobility $\mu$. 
An increase in $n$ typically reduces the asymmetry in the velocity of charge carriers about the chemical potential, causing a reduction in $S$.
Thus, typical metals exhibit a negligibly small $S$.

Recently, Kuroki and Arita theoretically showed that a metal with a peculiarly shaped band, referred to as the ``pudding mold'' type, which consists of a dispersive portion and a flat portion, should exhibit enhanced thermoelectric properties.\cite{JPSJ.76.083707}
When the chemical potential is located in the dispersive portion but near the flat portion, large asymmetry appears in the carrier velocity about the chemical potential, yielding large $S$ even for a metal with a large Fermi surface. 
Subsequently, the existence of such a ``pudding mold''  band was discovered in the thermoelectric oxide Na$_x$CoO$_2$ by angle-resolved photoemission spectroscopy. \cite{PhysRevLett.92.246403,PhysRevLett.95.146401,1367-2630-13-4-043021}
As predicted, 
Na$_x$CoO$_2$ exhibits large $S$ ($\simeq$ 100 $\mu$V/K) and low $\rho$ ($\simeq$ 200 $\mu\Omega$cm) at 300 K, \cite{PhysRevB.56.R12685,motohashi:1480} which yields a large PF of $\simeq$ 50 $\mu$W/cmK$^2$, comparable with that of a typical thermoelectric material Bi$_2$Te$_3$ (PF $\simeq$ 40 $\mu$W/cmK$^2$). \cite{Caillat19921121}
In addition, the thermoelectric properties of Na$_x$CoO$_2$ are further enhanced at elevated temperatures: $ZT$ reaches a value of 1 at 800 K. \cite{JJAP.40.4644}
In addition, enhanced thermoelectric properties have also been observed in a number of cobalt oxides \cite{funahashi:2385,shikano:1851} and rhodium oxides,  \cite{JPSJ.75.023704,PhysRevB.74.235110} where the ``pudding mold'' type band is also thought to play an important role. \cite{PhysRevB.78.115121,0953-8984-21-6-064223}

The title compound PtSb$_2$ crystallize in a cubic pyrite structure (space group Pa$\bar{3}$), which consists of edge-sheared PtSb$_6$ octahedra that are tilted to form diatomic molecules of [Sb$_2$]$^{4-}$. 
The filled $t_{2g}$ orbital of Pt$^{4+}$ (5d$^6$ low-spin state) is analogous to the Co$^{3+}$ (3d$^6$ low-spin state) of NaCoO$_2$, and accounts for the semiconducting nature of PtSb$_2$. 
Interestingly, band calculations for PtSb$_2$ suggest the existence of a ``pudding mold'' type valence band, \cite{PhysRev.138.A246}
suggesting that it may also exhibit enhanced thermoelectric properties.
In this Letter, we report the occurrence of a semiconductor-to-metal transition upon partial substitution of Ir for Pt in Pt$_{1-x}$Ir$_x$Sb$_2$ at $x$ $<$ 0.01. 
A large Seebeck coefficient emerges in the metallic state of doped Pt$_{1-x}$Ir$_x$Sb$_2$, 
resulting in a large PF of 43 $\mu$W/cmK$^2$ at 400 K.
Here we discuss whether the presence of a ``pudding mold'' type band is responsible for the enhanced thermoelectric properties of Ir-doped PtSb$_2$.

Polycrystalline samples of Pt$_{1-x}$Ir$_x$Sb$_2$ with $x$ = 0.0, 0.01, 0.03, and 0.1 were synthesized by a solid-state reaction in two steps. 
First, stoichiometric amount of starting materials Pt (99.99\%), Ir (99.99\%), and Sb (99.99\%) were mixed and ground. They were heated in an evacuated quartz tube at 1150$^\circ$C for 1 week. 
Then the product was powdered, pressed into pellets, and sintered at 1000$^\circ$C for 10 h. 
The obtained samples were characterized by powder X-ray diffraction (XRD) and confirmed to be a single phase of Pt$_{1-x}$Ir$_x$Sb$_2$. Thermoelectric properties, namely, electrical resistivity $\rho$, Seebeck coefficient $S$, and thermal conductivity $\kappa$, were measured using a Physical Property Measurement System (PPMS, Quantum Design) in the temperature range from 2 to 300 K. 
High-temperature $\rho$ and $S$ were measured using a thermoelectric measurement system (ZEM-3, ULVAC-RIKO) in the temperature range from 320 to 800 K.

\begin{figure}[t]
\begin{center}
\includegraphics[width=8cm]{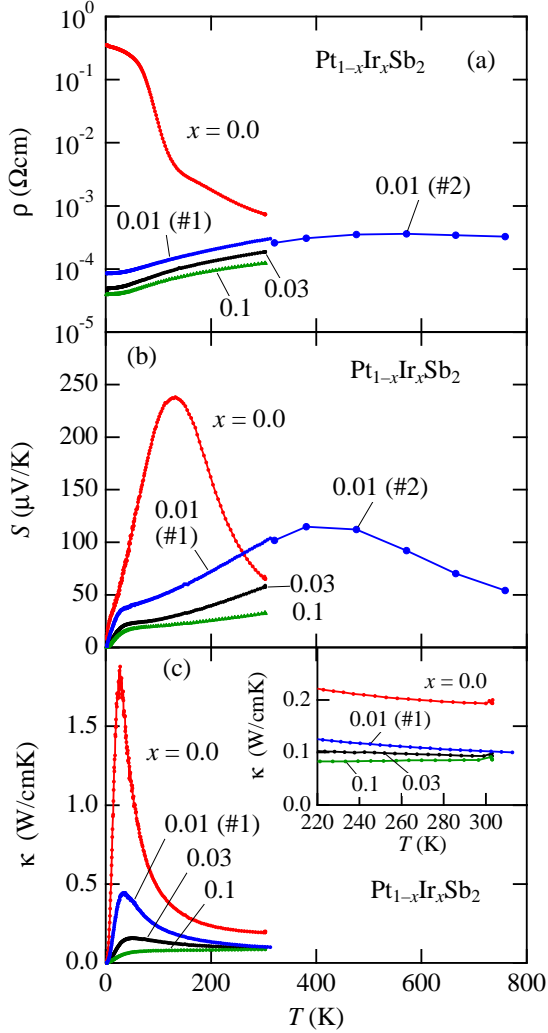}
\caption{\label{fig1}
(Color online) Temperature dependence of (a) electrical resistivity $\rho(T)$, (b) Seebeck coefficient $S(T)$, and (c) thermal conductivity $\kappa(T)$ of polycrystalline Pt$_{1-x}$Ir$_x$Sb$_2$ with $x$ = 0.0, 0.01, 0.03, and 0.1.
The inset of (c) shows $\kappa(T)$ data between 220 and 300 K.
}
\end{center}
\end{figure}

As shown in Fig.~1(a), non-doped PtSb$_2$ exhibits semiconducting behavior, consistent with the previous reports.  \cite{Johnston1965272}
The finite value of $\rho$ (= 0.35 $\Omega$cm) in the $T$ = 0 limit suggests the presence of extrinsic charge carriers for the PtSb$_2$ sample.
Then, both the magnitude and temperature dependence of $\rho$ changed abruptly by Ir doping from semiconducting to metallic.
The $\rho$ values on the order of 100 $\mu\Omega$cm and the positive temperature coefficient of $\rho$ suggest that a metallic state is realized for Pt$_{1-x}$Ir$_x$Sb$_2$ ($x$ = 0.01, 0.03, and 0.1). 
The Seebeck coefficient $S$ exhibits a positive value for all samples, as shown in Fig.~1(b), indicating that the majority of the charge carriers are holes. 
The $S$ value of the non-doped PtSb$_2$ exhibits a large maximum value of +250 $\mu$V/K at approximately 120 K, which is consistent with the literature data.\cite{Johnston1965272}
The temperature dependence of $S$ is changed abruptly by Ir doping. 
$S$ increases with temperature over the wide temperature range investigated in the present study for Pt$_{1-x}$Ir$_x$Sb$_2$ ($x$ = 0.01, 0.03, and 0.1). 
The $S$ value of the $x$ = 0.01 sample 
reaches a maximum of +112 $\mu$V/K at approximately 400 K.
In this way, metallic $\rho$ is compatible with a large $S$ for Pt$_{1-x}$Ir$_x$Sb$_2$.
The sample with $x$ = 0.01 exhibited the highest PF value of 
43 $\mu$W/cmK$^2$ at approximately 400 K,
as shown in Fig.~2(a).
This value is comparable to that for Bi$_2$Te$_3$ (40 $\mu$W/cmK$^2$). \cite{Caillat19921121}

\begin{figure}[t]
\begin{center}
\includegraphics[width=8cm]{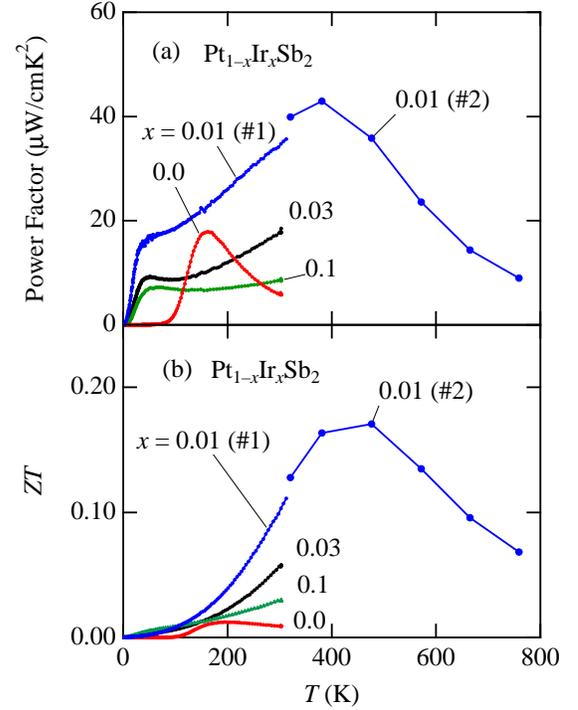}
\caption{\label{fig2}
(Color online) Temperature dependence of (a) power factor (PF) and (b) thermoelectric dimensionless figure of merit $ZT$ of polycrystalline Pt$_{1-x}$Ir$_x$Sb$_2$ with $x$ = 0.0, 0.01, 0.03, and 0.1.
We estimated $ZT$ values at $T$ $>$ 320 K using a value of $\kappa$ = 0.1 W/cmK at 300 K.
}
\end{center}
\end{figure}

Furthermore, we found that 
partial substitution of Sn for Sb results in similar behaviors. 
Pt(Sb$_{1-x}$Sn$_x$)$_2$ with $x$ = 0.01 exhibits metallic behavior as well as a large $S$, resulting in a PF of $\simeq$ 20 $\mu$W/cmK$^2$ at 300 K.

The origin of the enhanced PF of the Ir- and Sn-doped PtSb$_2$ is inferred from the previous band calculations. \cite{PhysRev.138.A246}
For PtSb$_2$, the calculations suggest that the valence band exhibits very shallow maxima on $\langle100\rangle$ axes and a minimum at the $\Gamma$ point with an energy difference of 0.011 eV, \cite{PhysRev.138.A246} which can be viewed as the flat portion of the ``pudding mold'' type band. 
Six hole pockets are expected on the $\langle100\rangle$ axes for very slightly doped PtSb$_2$ when the chemical potential lies between the band maxima and minima, as have been observed experimentally. \cite{PhysRevB.5.2175}
The pockets would fuse into a single (large) Fermi surface when the chemical potential is lowered across the band minima upon further doping. 
Such a topological change in the Fermi surface can be expected to occur at a carrier density of $n_{\rm c}$ = 3 $\times$ 10$^{19}$ cm$^{-3}$ according to the number of electronic states. \cite{PhysRev.138.A246}
For Pt$_{1-x}$Ir$_x$Sb$_2$ with $x$ = 0.01, we estimate the number of charge carriers $n$ to be 4.5 $\times$ 10$^{20}$ cm$^{-3}$, which is larger than $n_{\rm c}$,  by assuming that one hole is introduced per Ir atom because Ir has one less electron than Pt. 
Thus, the chemical potential is expected to lie at the dispersive portion of the band but not far below the flat portion. 
This results in a large asymmetry in the velocity of charge carriers about the chemical potential, which leads to a large $S$ along with metallic conductivity due to the large Fermi surface.
Further studies, such as photoemission spectroscopy, can be invaluable to further elucidate the causes for these enhanced properties.

Finally, we examine the thermoelectric efficiency of the doped Pt$_{1-x}$Ir$_x$Sb$_2$.
As shown in Fig.~1(c), the $\kappa$ values decrease dramatically with increasing Ir content $x$ and converge to $\kappa$ $\simeq$ 0.1 W/cmK at 300 K for the doped samples.
However, this value is approximately one order of magnitude larger than that for Bi$_2$Te$_3$. \cite{Caillat19921121}
This is the main reason for the suppressed value of the dimensionless figure of merit ($ZT = S^2T/\rho\kappa$). 
As shown in Fig.~2(b), the calculated dimensionless figure of merit $ZT$ for the sample with $x$ = 0.01 
reaches a maximum of 0.17 at approximately 480 K.
Here, we assumed that $\kappa$ is independent of temperature at $T$ $>$ 300 K and used a value of $\kappa$ = 0.1 W/cmK at 300 K to estimate $ZT$ at $T$ $>$ 300 K.
The sample shows a relatively high PF at approximately 400 K, lending support for further investigation into the present system to reduce thermal conductivity $\kappa$ and enhance $ZT$.
We estimated the electronic thermal conductivity values, $\kappa_e$, using the Wiedemann$-$Franz relation ($\kappa_e = L_0T/\rho$, where $L_0$ is the Lorenz number 2.44 $\times$ 10$^{-8}$ W$\Omega$/K$^2$) from the measured $\rho$ values. 
We found that $\kappa_e$ = 0.025 W/cmK for Pt$_{1-x}$Ir$_x$Sb$_2$ ($x$ = 0.01) at 300 K. 
The $\kappa$ value at 300 K can be then calculated as the sum of the electronic and lattice components.
This lattice part, $\kappa_{\rm{lattice}}$ $\simeq$ 0.075 W/cmK, should be reduced further to realize improved thermoelectric properties in Ir-doped PtSb$_2$.

In summary, polycrystalline samples of Pt$_{1-x}$Ir$_x$Sb$_2$ with $x$ = 0.0, 0.01, 0.03, and 0.1 were prepared and their thermoelectric properties were investigated at 2$-$800 K. 
The doped samples exhibited metallic conductivity and a large Seebeck coefficient, which resulted in an enhanced power factor of approximately 43 $\mu$W/cmK$^2$ at 400 K for Pt$_{1-x}$Ir$_x$Sb$_2$ at $x$ = 0.01. 
The abrupt changes in $\rho$ and $S$ upon doping at $x$ $<$ 0.01, together with the band calculation, \cite{PhysRev.138.A246} suggests that a ``pudding mold'' type band with shallow energy minima and maxima is present in PtSb$_2$ and plays an important role in the observed metallic conductivity and large Seebeck coefficient, which are a key features for practical thermoelectric materials.

Part of this work was performed at the Advanced Science Research Center, Okayama University.
This work was partially supported by a Grants-in-Aid for Young Scientists (B) (23740274) from the Japan Society of the Promotion of Science (JSPS) and the Funding Program for World-Leading Innovative R\&D on Science and Technology (FIRST Program) from JSPS.

\end{document}